\documentclass[showpacs,preprintnumbers,prb,twocolumn,floatfix]{revtex4}

\usepackage{graphicx}
\usepackage{dcolumn}
\usepackage{bm}
\usepackage{pstricks}
\usepackage{amssymb}
\usepackage{amsmath}

\begin{document}

\title{Inducing energy gaps in graphene monolayer and  bilayer}
\author{R.~M. Ribeiro  and  N.~M.~R. Peres}
\affiliation{Center of Physics  and  Department of Physics,
             University of Minho, PT-4710-057, Braga, Portugal}
\email{ricardo@fisica.uminho.pt}

\author{J. Coutinho}
\affiliation{Departamento de F\'{\i}sica, Universidade de Aveiro,
             Campus Santiago, 3810-193 Aveiro, Portugal}

\author{P. R. Briddon}
\affiliation{School of Natural Sciences, Newcastle University,
             Newcastle upon Tyne NE1 7RU, United Kingdom}
\begin{abstract}
In this paper we study the formation of energy gaps in the spectrum of graphene and its
bilayer when both these materials are covered with water and ammonia molecules.
The energy gaps obtained are within the range 20-30 {\ttfamily meV}, values
compatible to those found in experimental studies of graphene bilayer.
We further show that the binding energies are large enough for the adsorption of the molecules
to be maintained even at room temperature.
\end{abstract}

\pacs{71.15.-m,71.15.Mb,71.20.Tx,73.22.-f,}

\maketitle

\section{Introduction}

Graphene is a two dimensional system made solely of Carbon atoms arranged
in a honeycomb lattice. Since its isolation\cite{novo1,pnas} 
in the late 2004, it
has attracted a great deal of attention due to its unconventional physical
properties. Together with this wonderful system 
it was also possible to produce its layered relatives,
like the graphene bilayer, trilayer, etc. 

The physical properties of single layer graphene are governed by the
elementary excitations around the Fermi surface, which in graphene is made of
only
six points in the Brillouin zone, located at momenta values denominated
$\mathbf{K}$ and $\mathbf{K'}$ points \cite{rmp}. These elementary excitation are 
massless Dirac particles, described by the $2+1$ Dirac equation. The unusual
properties of graphene are therefore a consequence of the fact that the emergent
physical theory of this system is not the usual Schr\"odinger equation but
instead a relativistic problem. For the layered systems the situation is also
equally interesting. For example for the bilayer case, although the low
energy spectrum is parabolic, the effective theory is that of massive chiral
particles\cite{falko,johan1,johan2}
and again not the usual  Schr\"odinger equation. 

Both graphene and its bilayer are expected to integrate the next
generation of nanoelectronic devices\cite{johan3}. However, both graphene
and its bilayer do not present an energy  gap in the spectrum, preventing
large on-off signal ratios in possible electronic devices. In the case of 
graphene, there is also a finite conductivity
 value\cite{novo1,pnas,Peres06} at the neutrality point,
which prevents the pinch-off of the field effect transistor. Clearly, what
is necessary for having full working devices made of graphene is 
an energy gap in the spectrum, as in the usual semiconductors devices. 

There are at the moment two different proposals for producing 
energy gaps in the spectrum of graphene. The first is by producing graphene
nanoribbons \cite{Louie06,Duan07,Dai08}. 
The theoretical studies \cite{Louie06,Duan07} show that depending on the type
of terminations a graphene ribbon may have, the mechanisms for the opening 
of a gap will be different. For zig-zag nanoribbons, 
the existence of edge
states\cite{Fujita,Nakada,eduardo08} would prevent the existence
of an energy gap. However, the system finds a way out by inducing
magnetization\cite{Louie06} at the edges and opening a gap in the 
spectrum. The same happens in a graphene bilayer with zig-zag
edges \cite{eduardo08,eduardoEDGE}. For nanoribbons with armchair
edges, the gaps are formed due to quantum confinement, and the value
of the gap depends on the width of the ribbon. For 
ribbons smaller than 10~nm on/off ratios as large as
10$^6$ have been demonstrated \cite{Dai08}.

Another method of producing a gap in the spectrum of graphene is by
depositing graphene on top a boron nitride (BN)
\cite{Giovannetti07}. This material is a band gap insulator with a
boron to nitrogen distance of the order of 1.45~\AA \cite{Zupan71},
(in graphene the carbon-carbon distance is 1.42~\AA) and a gap of
the order of 4~{\ttfamily eV}. It was shown that in the most
stable configuration, where a carbon  is on top of a boron and the
other carbon in the unit cell is centered above a BN ring, the value
of the induced gap is of the order of 53~m{\ttfamily eV}. Depositing
graphene on a metal surface with a BN buffer layer leads to $n-$doped
graphene with an energy gap of 0.5 {\ttfamily eV}\cite{Lu07}.

The existence of energy gaps in the spectrum of graphene
also prevents the occurrence of the Klein paradox
\cite{NaturePhys,Bai07}, which would be another limiting source to the 
effective pinch off of the field effect transistor.

The situation in bilayer graphene is also being unveiled experimentally.
It has already been demonstrated that it is possible to
open a gap in the spectrum of the bilayer
when this material is deposited on top of silicon carbide
(SiC) and its exposed face is covered with Potassium atoms \cite{Rothenberg}.
The value of the gap is, in this case, connected to the amount of doping
the potassium atoms induce. Using a similar method, it was experimentally demonstrated
the opening of a gap in a bilayer deposited on top of silicon oxide, by
covering its exposed face with ammonia molecules \cite{eduardogeim}.
In both these methods, the value of the gap and the amount of doping are
interconnected. Recently\cite{morpurgo}, a device with both a top- and a
back- gate setups became available, which allows to control the value of the
gap in the bilayer spectrum and the amount of charge carrier independently.

>From the above experimental studies  it is clear that covering the
bilayer system with some atomic species will lead to a gap in the spectrum.
These results motivated first principle studies\cite{Katsnelson08}
of gap formation in graphene bilayer leading to calculated gaps
in the interval 0.64-3 {\ttfamily eV}.

In this paper we  study the effect of covering both graphene and its
bilayer with water and ammonia, showing that in both cases this covering
leads to a gap in the spectrum, albeit smaller than those reported in
Ref. [\onlinecite{Katsnelson08}]. Our results for the induced gaps in the 
graphene bilayer are compatible with those determined for this system
by measuring the cyclotron masses of the charge carriers \cite{eduardogeim}.

\section{Theoretical Method}
We are interested in describing the effect that covering graphene and its bilayer
with water and ammonia has in their electronic spectra and in particular in the
appearance of an energy gap in the spectrum.  There is strong experimental evidence for 
the presence of water in graphene\cite{Moser}, and there are working graphene devices  based 
on covering one of the surfaces of the graphene bilayer with ammonia\cite{eduardogeim}. Therefore our motivation
is to understand from  first principles how these two types of molecules change the
electronic spectrum near the Dirac point.

The method we use in our investigation rests upon the
local density approximation (LDA). In particular, the calculations were performed using the density functional code {\sc aimpro}\cite{Jones1998,Briddon2000}. In what follows we give the relevant 
details of our calculations.

The Brillouin-zone (BZ) was sampled for integrations according to the scheme proposed by Monkhorst-Pack\cite{Monkhorst1968}. 
A grid of $8\times4\times1$~$\mathbf{k}$-points was generated and folded according to the symmetry of the 
BZ. An increase in the number of points did not result in a significant total energy change.

We use pseudopotentials to describe the ion cores.
Lower states (core states) are accounted for by using the dual-space separable pseudopotentials by Hartwigsen, Goedecker, and Hutter\cite{Hartwigsen1998}.
The valence states are expanded over a set of $s$-, $p$-, and $d$-like Cartesian-Gaussian Bloch atom-centered functions,
and Fermi-filled using a value of $k_B T =0.01$~{\ttfamily eV}.
Kohn-Sham states are expressed as linear combinations of these basis functions.
The basis functions were optimized for graphene, water and ammonia separately.
The calculation of potential terms is more efficient in reciprocal space, and therefore the electron density is Fourier transformed by using plane waves with kinetic energy of up to 300~Ry.

Graphene was modeled in a slab geometry by including a vacuum region in a supercell containing 8 carbon atoms.
Different sizes were also tested, as well as several water or ammonia molecules concentrations per supercell,
leading to very different molecular surface densities but to only small changes in the gap values.
In the normal direction (z-direction),  the vacuum separating repeating slabs has more than 30~\AA.
The size of the supercell in the $z$-direction was optimized to make sure there was no interaction between repeating slabs.

In the calculations, all the atoms were fully relaxed to their equilibrium positions. That is to say that
the final structure of the system is that obtained by energy minimization and is not imposed externally.
Consequently all the distances found below are a consequence of energy minimization of the full structure.

The electronic band structure for the path $\mathbf{K-\Gamma-M}$ was calculated for all the structures. Since the supercell is much larger than the primitive cell, the bands appear folded in the graphs.
The electronic density of states (DoS) was calculated by sampling 12800 points in the BZ, and broadened with a 0.1~
{\ttfamily eV}  Gaussian width.

The binding energies were calculated by subtracting the total energies of the graphene sample and the molecule calculated separately from the total energy of the graphene plus molecule.

\section{Results}
\subsection{Structure}

\begin{figure}
 \includegraphics[scale=0.09,bb=-124 -104 1087 925]{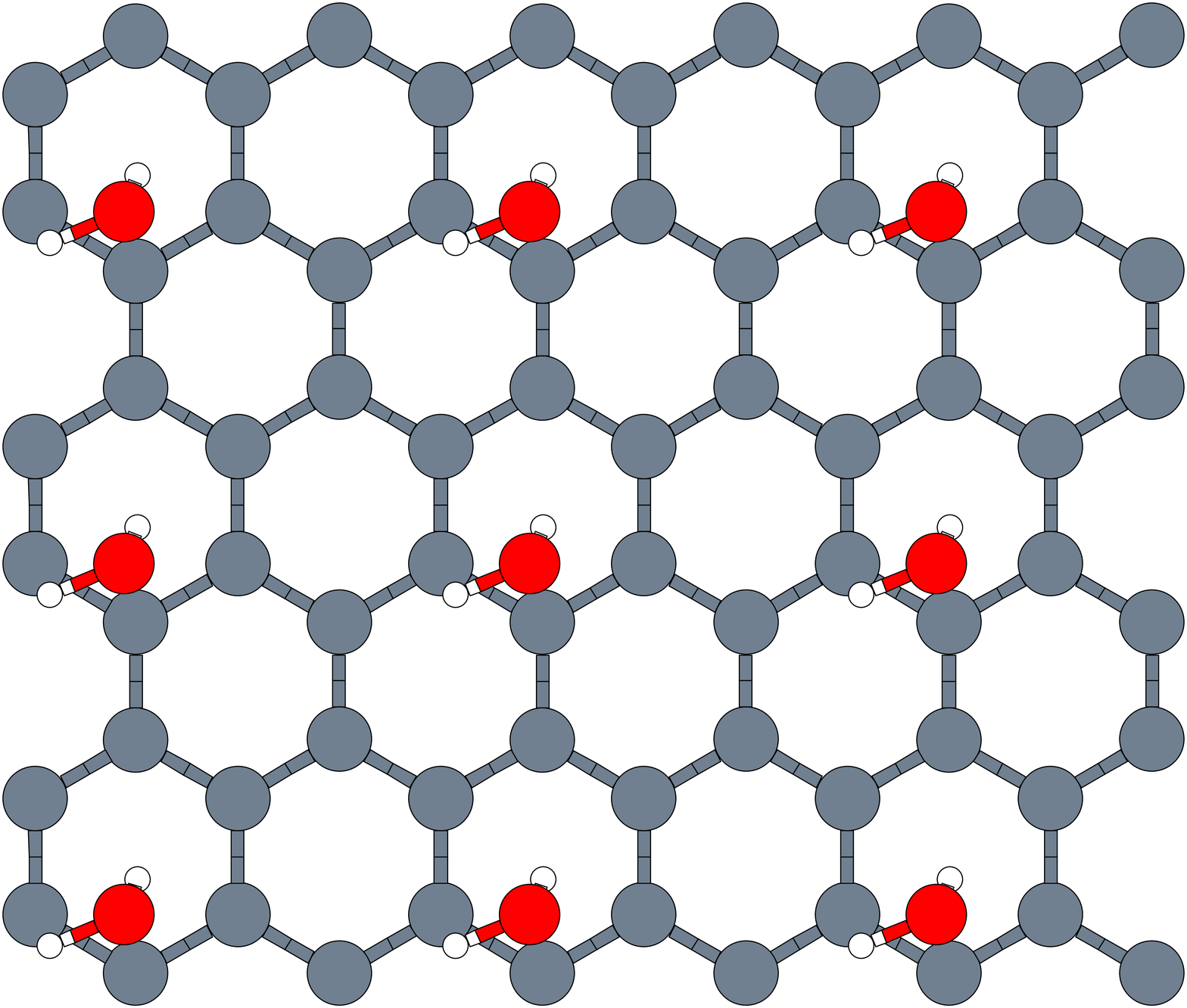}\hspace*{0.5cm}
 \includegraphics[scale=0.09,bb=-125 326 1088 635]{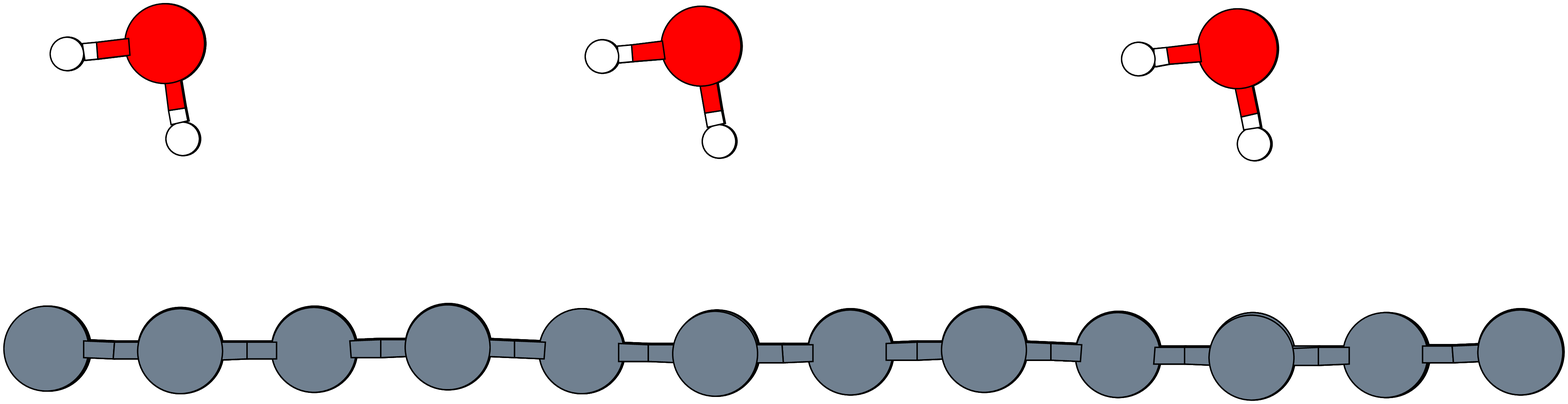}

 \caption{(Color online) Water on the top of graphene: top and side views.}
 \label{fig:gw-8c}
\end{figure}

Figure \ref{fig:gw-8c} shows the relaxed water molecule on the top of the graphene.
In this calculation, water concentration is very low (one water molecule per 8 carbon atoms,
or a density of about 6~nm$^{-2}$)
and this results in low interaction between the water molecules, meaning that each water molecule
acts like an independent scattering center.
One of the hydrogen atoms binds to a carbon atom, and the carbon atoms move 
from their initial positions, specially in the $z$-direction (as much as 0.05~\AA).
The presence of water therefore breaks the A-B symmetry of the graphene lattice structure.  This, as we shall see, will affect the electronic states (see Section \ref{electronic}), specially close to the
Dirac point. The same happens when we add ammonia.

\begin{figure}
 \includegraphics[scale=0.09,bb=-124 -104 1087 925]{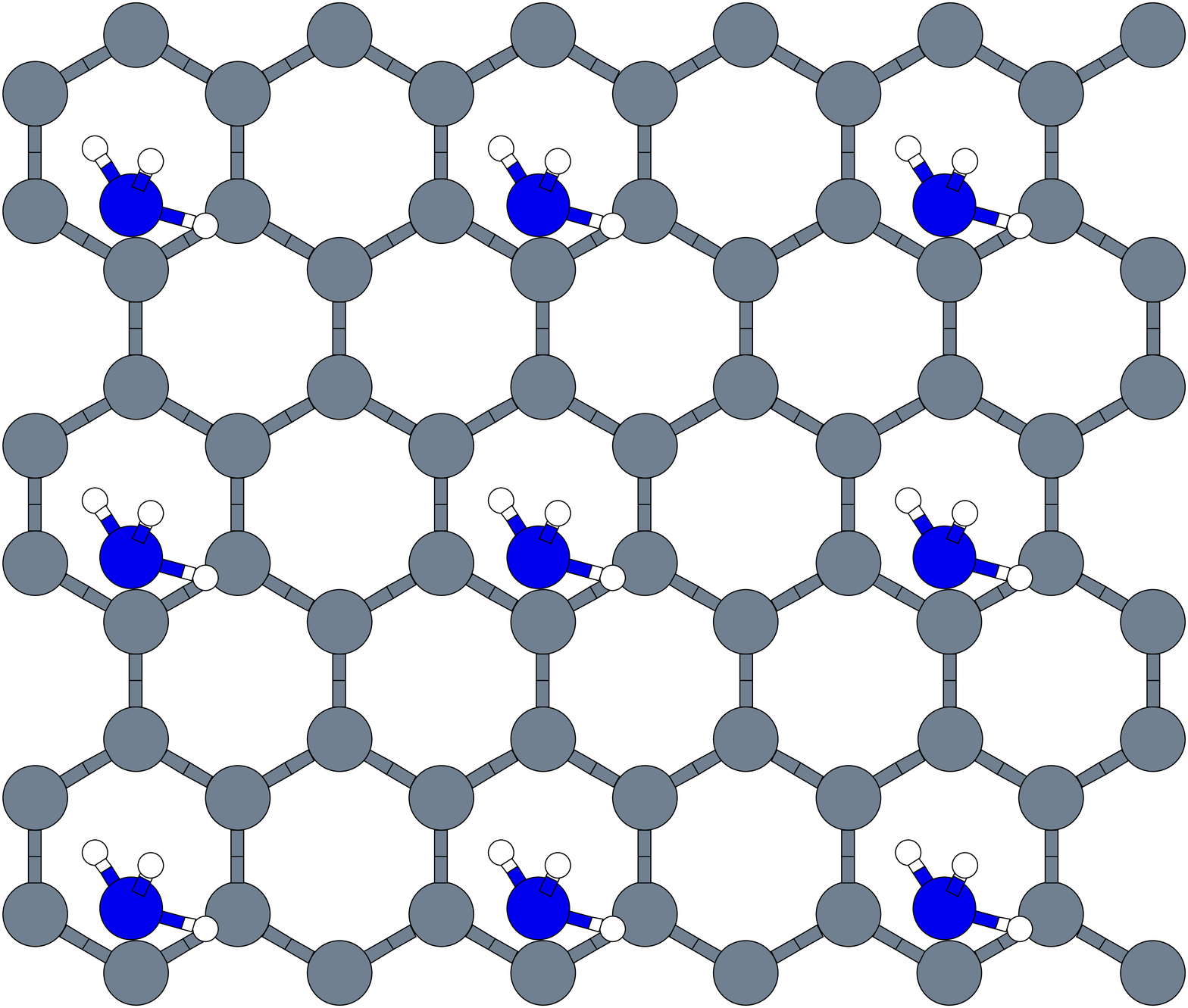}\hspace*{0.5cm}
 \includegraphics[scale=0.09,bb=-125 326 1088 635]{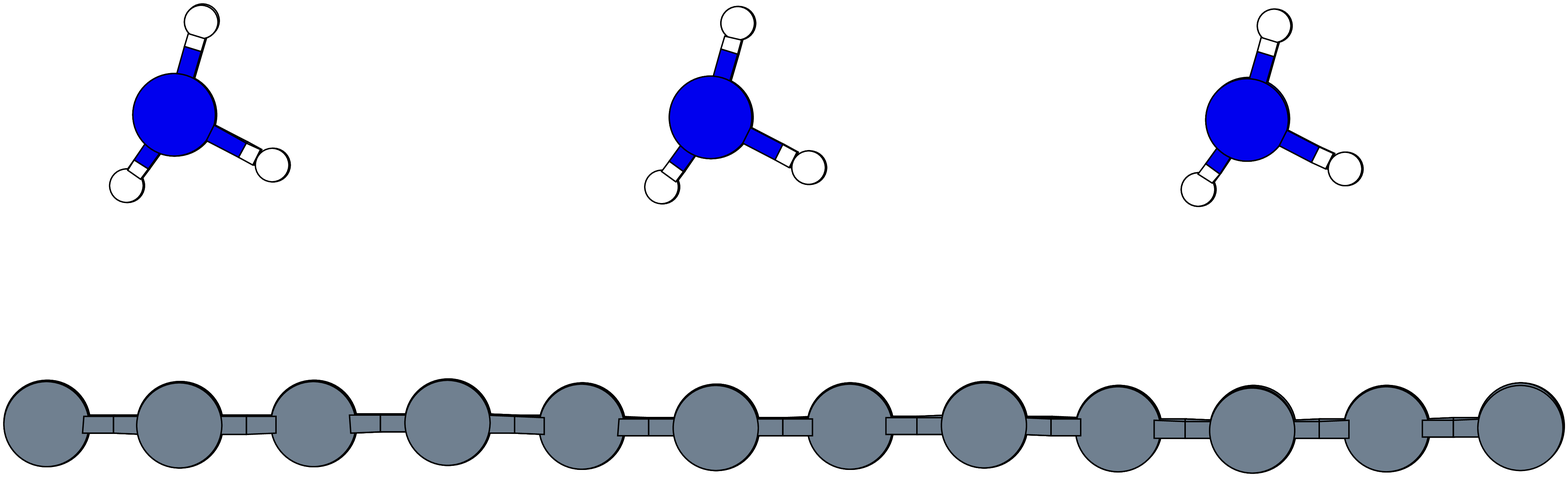}

 \caption{(Color online) Ammonia on the top of graphene: top and side views.}
 \label{fig:ga-8c}
\end{figure}

The ammonia in the top of the graphene (figure \ref{fig:ga-8c}) has a higher binding energy 
(see table \ref{table:energies}) since it binds through two hydrogen atoms, and not through one, as the water molecule.
Table \ref{table:distances} shows the distances between the heaviest atom of the water and the ammonia molecules and the closest carbon atom.
We choose the position of the heavier atoms of the molecules as a reference because their position
 can usually
be measured using techniques such as low-energy electron diffraction (LEED), X-ray photoemission
spectroscopy (XPS), ultraviolet photoemission spectroscopy (UPS), and electron-energy-loss
spectroscopy (EELS)\cite{Taylor1997}.
Our results show a much higher binding energy than the ones obtained by other authors\cite{Leenaerts2008}. We believe this is because those authors impose restrictions to the rotation
of the molecules on the top of graphene; we observed that our molecules never relaxed to those 
positions, but rotate instead to more favorable orientations.

When a second layer of graphene is added, a bilayer is obtained.
The usual stacking of the graphene bilayer is the Bernal staking, although there are situations where
rotational defects may also be present\cite{JLS}. 
In the Bernal stacking two of the carbon atoms, each belonging to different layers, are positioned exactly
on top of each other. For these atoms we have found that they move  away from the plane as much as 0.004~\AA.
The distance between layers (measured as the distance between two carbon atoms that are superimposed in the Bernal stacking) is 2.6~\AA. This is lower than the interplain distance on graphite. 
We believe this is due to the absence of layers on both sides of the bilayer.

The bilayer has a binding energy per carbon atom of 0.21~{\ttfamily eV}, and both the water and the ammonia molecules are less
bound to the bilayer than to graphene.
This is because in the bilayer the carbon atoms near the molecules also have to share electrons with the underlying  carbon layer. In fact, the hydrogen of the molecule that is closer to the surface binds to a carbon atom that has no other carbon atom below.
In every case the binding energy is high and so desorption by thermal evaporation at room temperature is not expected.
This is consistent with the experimental finding that graphene usually has adsorbed water\cite{Moser}.

\begin{figure}
 \includegraphics[scale=0.09,bb=-124 -104 1087 925]{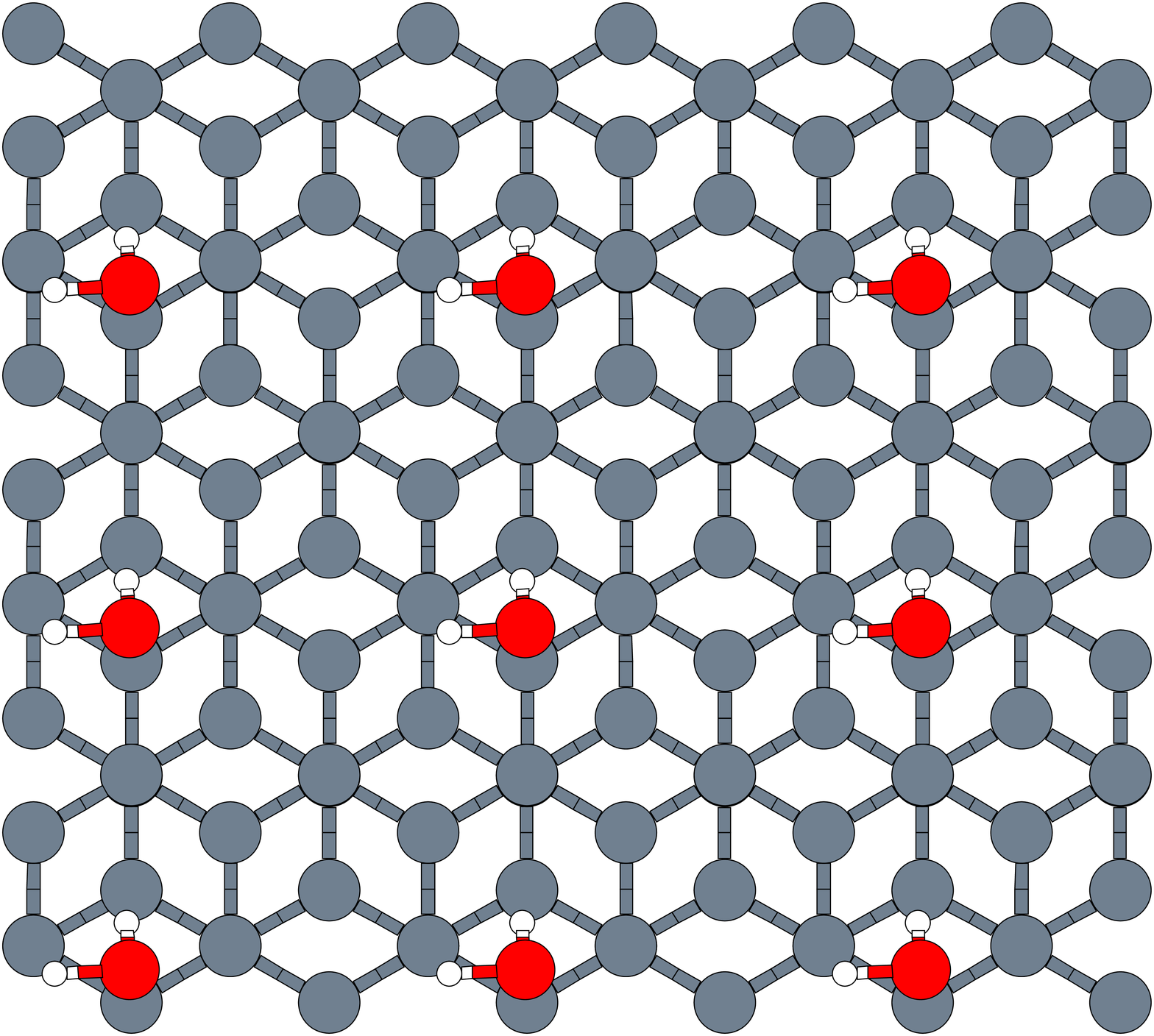}\hspace*{0.5cm}
 \includegraphics[scale=0.09,bb=-125 326 1088 635]{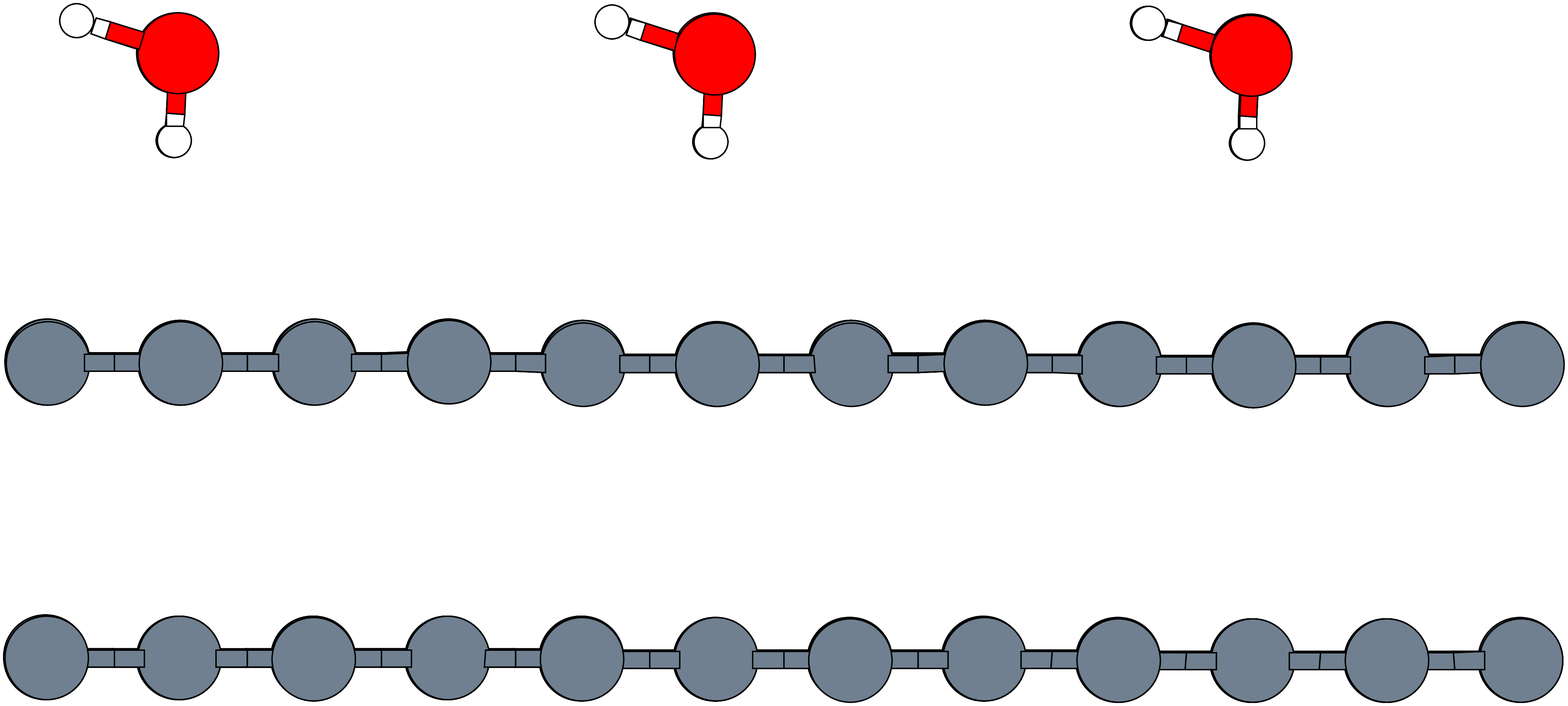}
 \caption{(Color online) Water on the top of bilayer graphene: top and side views.}
 \label{fig:2gw-8c}
\end{figure}
\begin{figure}
 \includegraphics[scale=0.09,bb=-124 -104 1087 925]{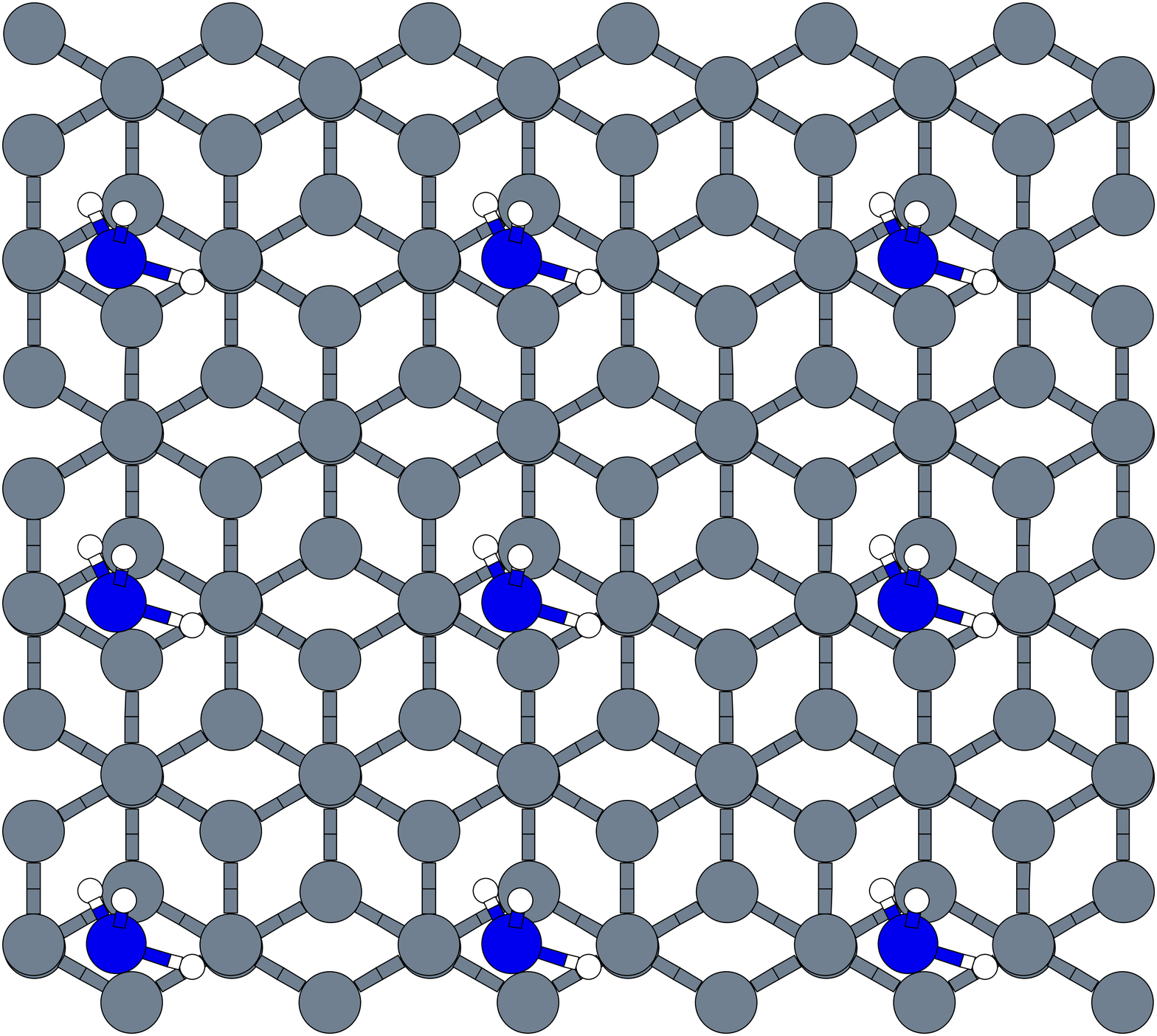}\hspace*{0.5cm}
 \includegraphics[scale=0.09,bb=-125 326 1088 635]{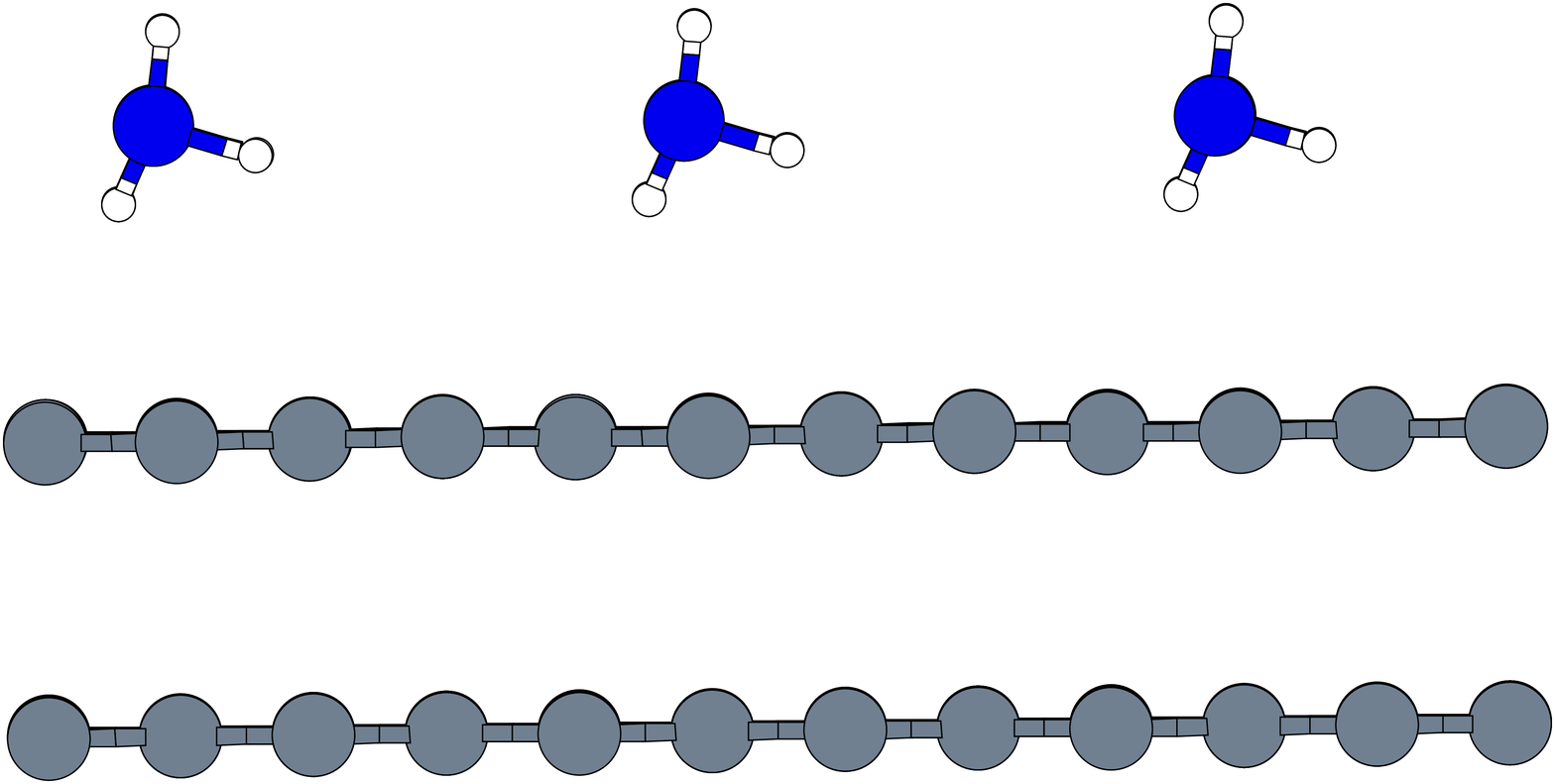}
 \caption{(Color online) Ammonia on the top of bilayer graphene: top and side views.}
 \label{fig:2ga-8c}
\end{figure}

\begin{table}
\caption{Distances between oxygen and nitrogen and the closest carbon atom of graphene}
 \begin{tabular}{llr}
  \hline
  System\hspace*{1cm}	& Atoms	\hspace*{1cm}	& Distance (\AA{})	\\
  \hline
  \hline
  Graphene	& O---C		& 2.79			\\
  Graphene	& N---C		& 2.91			\\
  Bilayer	& O---C		& 2.83			\\
  Bilayer	& N---C		& 2.97			\\
  \hline
 \end{tabular} 
\label{table:distances}
\end{table}

\begin{table}
\caption{Binding energies for the adsorbed molecules of ammonia and water (per molecule), 
and for the bilayer (per carbon atom).}
 \begin{tabular}{lr}
  \hline
  System\hspace*{1cm}	&  Binding energy ({\ttfamily eV})	\\
  \hline
  \hline
  Graphene + H$_2$O	& 1.94	\\
  Graphene + NH$_3$	& 3.58	\\
  Bilayer		& 0.21	\\
  Bilayer + H$_2$O	& 1.30	\\
  Bilayer + NH$_3$	& 2.86	\\
  \hline
 \end{tabular} 
\label{table:energies}
\end{table}

We also have done calculations for higher molecular concentrations.
In these cases, there is strong interaction between the molecules that cause their re-orientation, in such a way as to align some of the hydrogen atoms to the oxygen/nitrogen in order to make hydrogen bonds.
For water, two molecules per 8 carbon atoms covers the graphene surface;  one molecule more and a second water layer starts to form.
For ammonia, two molecules per 8 carbon atoms is enough for a second layer to be formed.

\subsection{Electronic structure}
\label{electronic}
\begin{figure}
\begin{tabular}{cc}
 a)\includegraphics[scale=0.3,bb=50 50 410 302]{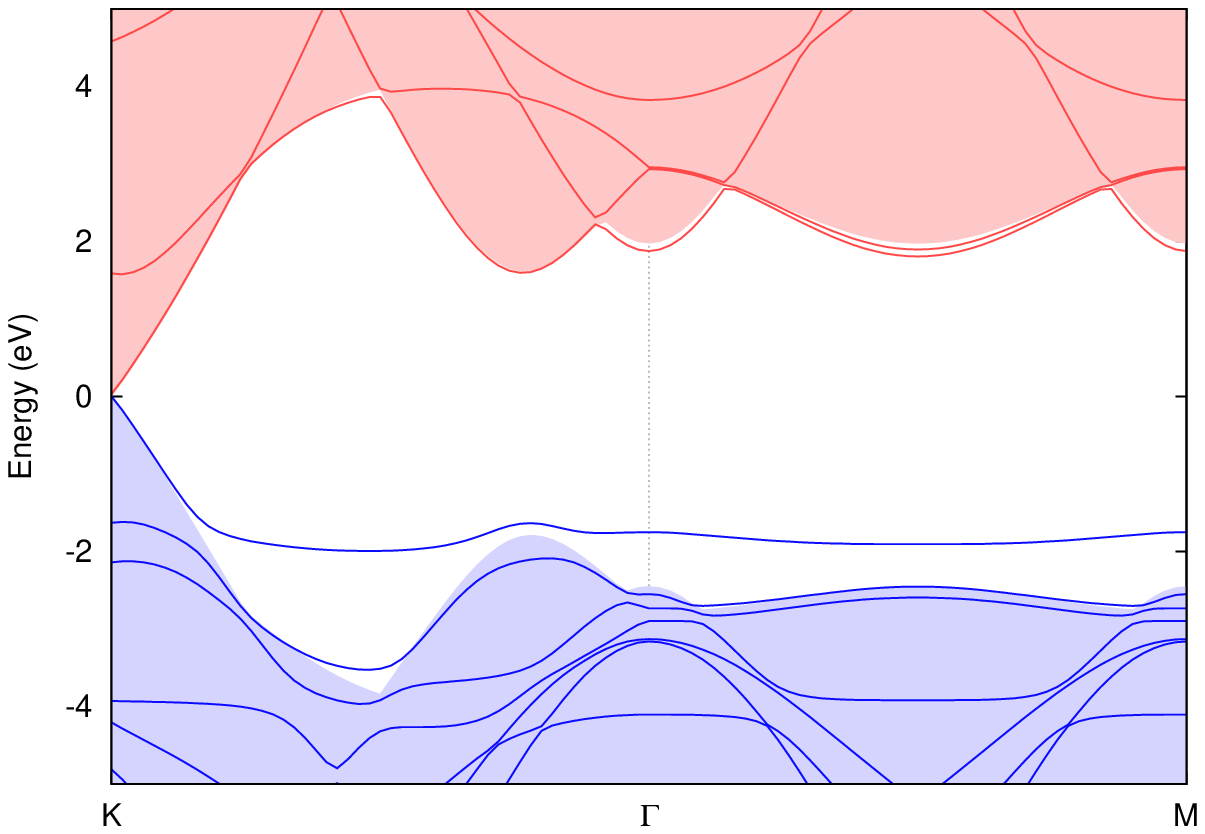}&
 c)\includegraphics[scale=0.3,bb=50 50 410 302]{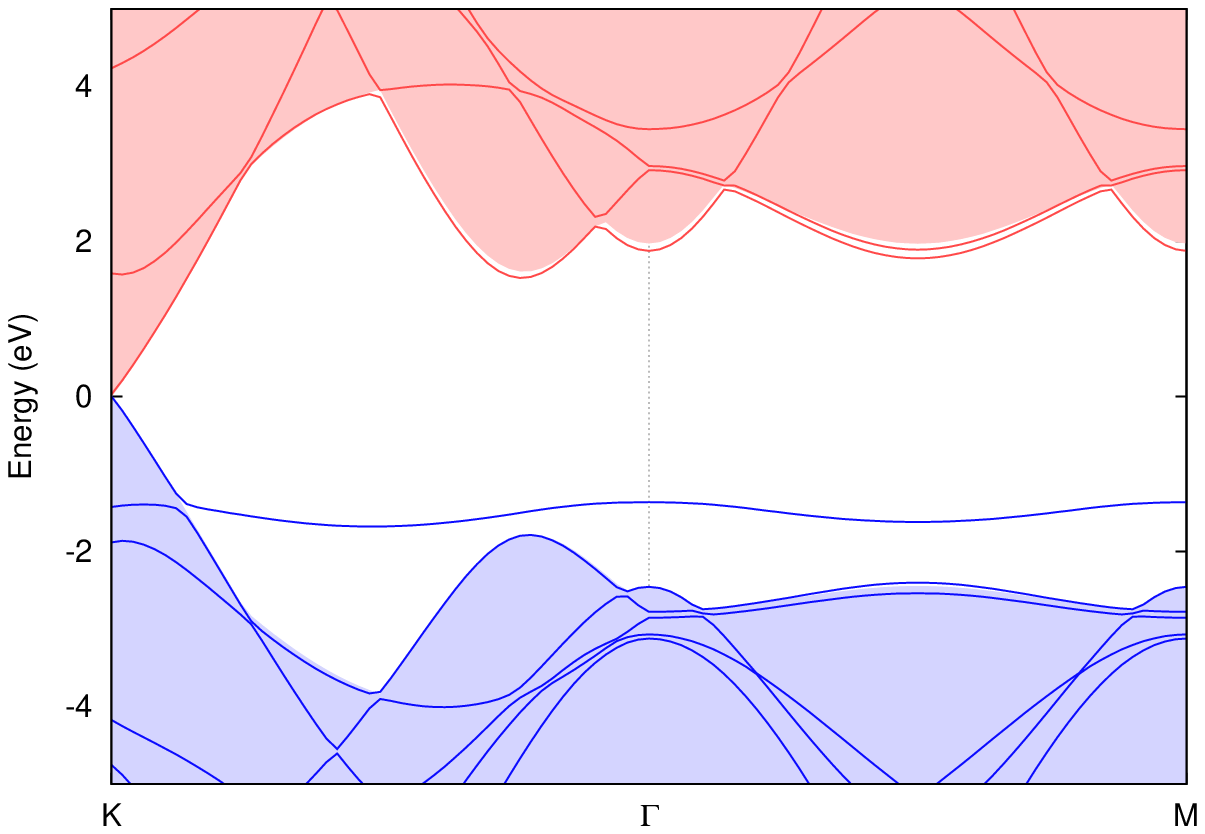}\\
 b)\includegraphics[scale=0.5,bb=50 50 194 150]{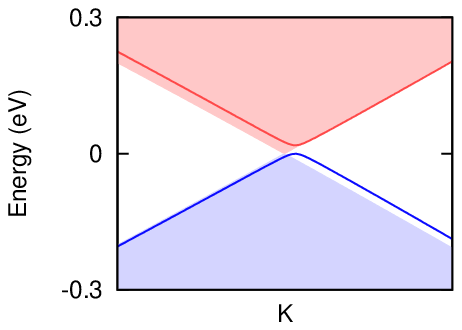}&
 d)\includegraphics[scale=0.5,bb=50 50 194 150]{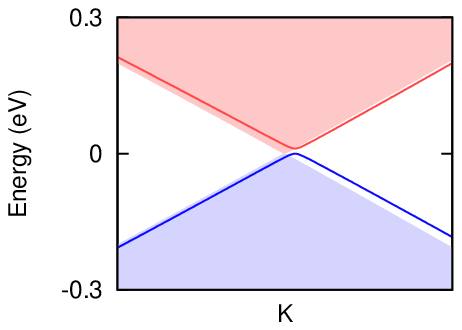}\\
\end{tabular}
\caption{(Color online) Electronic band structure for the system a) graphene plus water, b) detail near the Dirac point, c) graphene plus ammonia, d) detail near the Dirac point. The clean graphene band structure appears as a shadow in the background.}
\label{fig:band-g}
\end{figure}

The electronic band structure for the graphene with water and ammonia molecules on the surface is shown in figure \ref{fig:band-g}. Since the we are using a supercell, the Brillouin zone is
different from the standard one, and that is reflected on the shape of the bands. 
But the absolute location of points \textbf{K} and \textbf{M} are the same.
In both cases, a localized band appears distinctly above the valence band of the graphene.
It corresponds to the HOMO level of the molecules, which is clearly below the Fermi level.
The LUMO of the molecules is very much above the Fermi level.
This means that the charge transfer in the interaction between the molecules and the surface of graphene
will be small, and the interaction must be of dipole or Van-der-Waals type.

The bands near the Dirac point can be seen in detail in figure \ref{fig:band-g} and have a similar form for both types of molecular coverage of the carbon layers.
A gap opens and it is shifted to one side of the \textbf{K} point.
This happens because there is a symmetry lost due to the displacement of the carbon atoms interacting with the molecules.

Table \ref{tab:gap} shows the calculated values for the band gap.
It is well known that local density functional theory, while expected to give a realistic account of the ground state structure, it only gives an approximate description of unoccupied states in a system.
In particular the energy gap is underestimated, being typically 60\% off the real value.
This should be taken into account when analyzing the bands and the gap values in detail.
In principle a gap in the graphene spectrum is not expected, except when the A-B symmetry of the lattice
in broken. As our calculations show, there is a vertical distortion of the graphene lattice which leads
precisely to the breaking of the A-B symmetry and therefore to the opening of a gap. This breaking of the
symmetry lattice translates in a spectrum that is hyperbolic near the Dirac point, where the bands
are still linear for finite energies, except close to the Dirac point, where they become parabolic in shape.

\begin{figure}
\begin{tabular}{cc}
 a)\includegraphics[scale=0.3,bb=50 50 410 302]{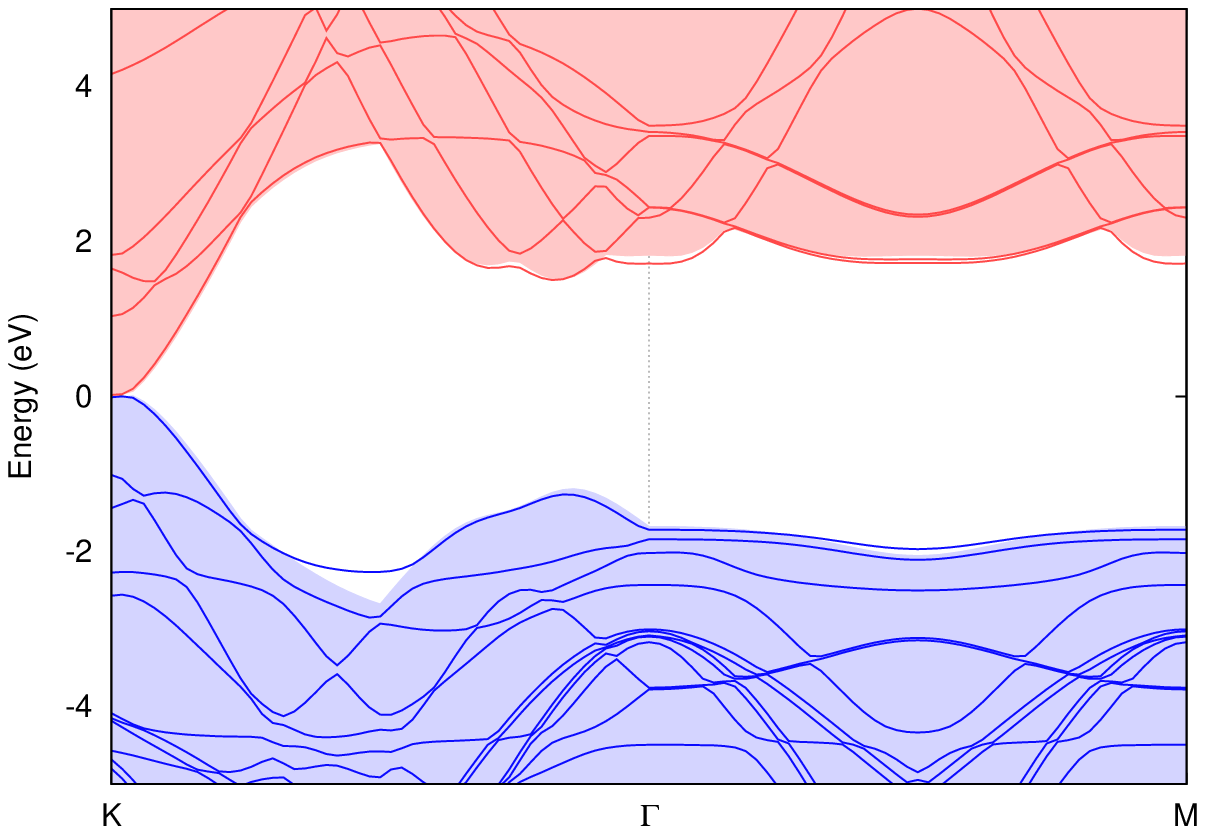}&
 c)\includegraphics[scale=0.3,bb=50 50 410 302]{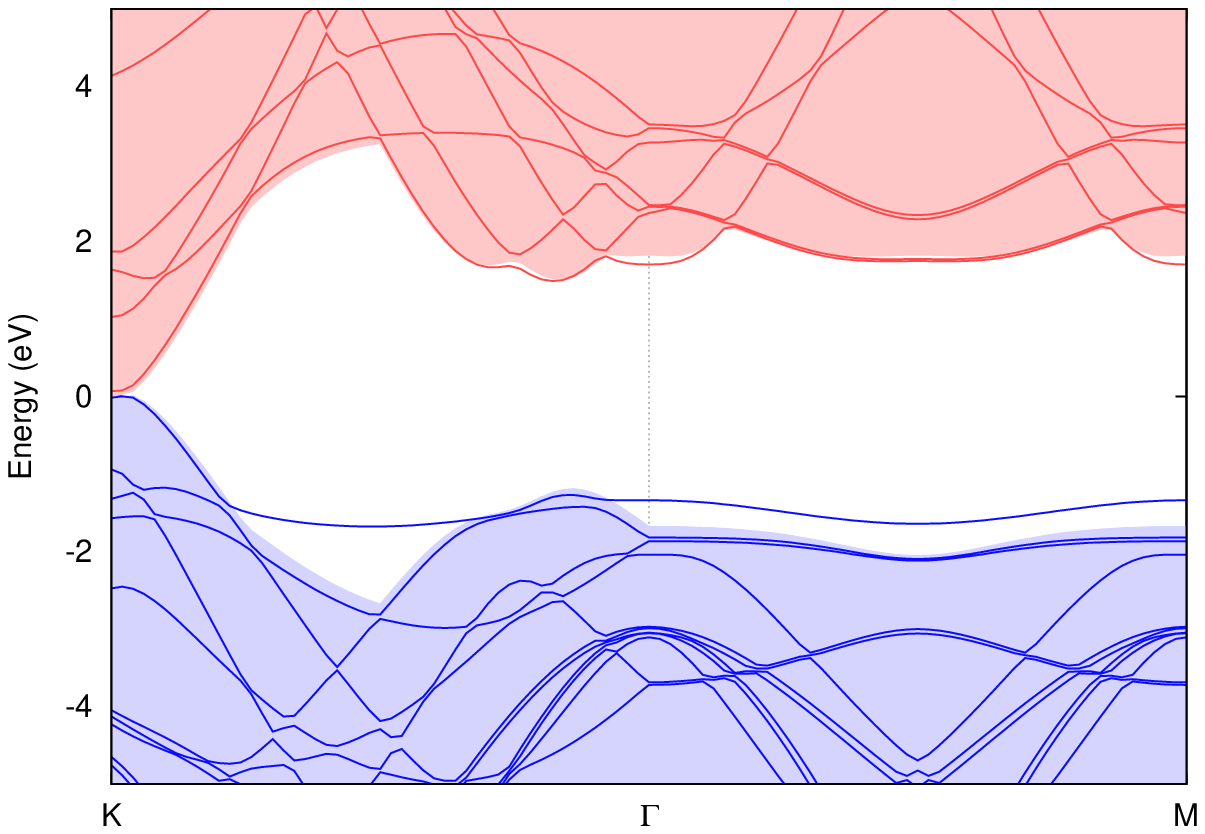}\\
 b)\includegraphics[scale=0.5,bb=50 50 194 150]{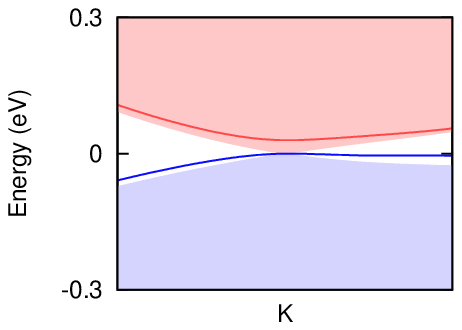}&
 d)\includegraphics[scale=0.5,bb=50 50 194 150]{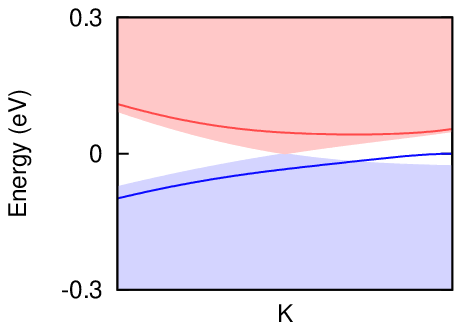}\\
\end{tabular}
\caption{(Color online)  Electronic band structure for the system a) bilayer plus water, b) detail near the Dirac point, c)  bilayer plus ammonia, d) detail near the Dirac point. The clean bilayer graphene band structure appears as a shadow in the background.}
 \label{fig:band-2g}
\end{figure}

The electronic band structure for the bilayer with water and ammonia molecules on the surface are shown in figure \ref{fig:band-2g}.
Since in our calculations we 
have the two layers much closer than in graphite, which is the distance
usually considered in other calculations, it is natural that we
have the second band at higher energies than other authors, due to an increase of the tight-binding
parameters in the direction perpendicular to the graphene layers.
In the case of water on the top of the bilayer there is no flat band above the bilayer valence band, while there is one in the case of ammonia.

As in graphene, covering the bilayer with either water or ammonia leads to the
opening of gap which  is shifted to one side of the \textbf{K} point, which is consistent with the results obtained by Guinea~\textit{et al.}\cite{Guinea2006}.
These authors found that the gap in bilayer graphene does not reach a minimum at the \textbf{K} point due to a ''Mexican hat'' dispersion at low energies.
Although the resolution of our figures is not enough to see ''Mexican hat'', its presence in the
band structure was verified by the authors.
The gaps (see Table \ref{tab:gap}) that the molecules generate on the bilayer are much larger than the ones on  graphene.

It is known that the unbiased bilayer has parabolic bands at the Dirac point, with the valence and the conducting band touching each other.
The presence of the molecules opens up a gap in the spectrum, but while in the case of water the gap is very near the Dirac point, in the case of ammonia the gap is indirect and far from the \textbf{K} point. This last result is a direct
consequence of the type of molecule used and can not be obtained within tight binding calculations.

It is known\cite{eduardogeim} that a difference in potential along the $z$-direction induces a gap in the electronic band structure for the bilayer.
The water and the ammonia molecules have both a electric dipole which will produce an electric field on the bilayer.
In this case, the difference in electric potential between the layers is 0.198~V for the system with ammonia and 0.027~V for the system with water.

The gap we obtained with ammonia is consistent with the gap calculated using tight binding and a selfconsistent determination of the electric potential\cite{eduardogeim}.
Yet, our calculations lead to a electric potential between graphene layers interacting with the ammonia molecule of 0.198~V.
This value, when included in the equation\cite{eduardogeim} 
\begin{equation}
\Delta_g = [e^2V^2t_{\bot}^2/(t_{\bot}^2 + e^2V^2)]^{1/2} 
\label{gap}
\end{equation}
gives a much larger gap.
The discrepancy may be explained by the fact that the expression is not valid for indirect gaps, and it does not include the distortion of the bilayer lattice.
The gap for the bilayer with water estimated from the electric potential between the layers and using Eq. (\ref{gap})
(based on a tight binding model) gives consistent results with those determined from the band structure calculation.

In our case, besides the electric potential induced by the molecule, there is also a distortion in lattice both in the case of graphene and its bilayer.
Our calculations show that the distortion of the lattice in the bilayer 
tends to counteract the molecule's dipole, but still leading to a band gap.

\begin{table}
\caption{Calculated band gaps for the systems considered. The gap shown for the bilayer+NH$_3$ is indirect.}
 \begin{tabular}{lr}
\hline
    System\hspace*{1cm}	& Band gap ({\ttfamily meV})\\
\hline
\hline
Graphene + H$_2$O	&   18	\\
Graphene + NH$_3$	&   11	\\
Bilayer + H$_2$O	&   30	\\
Bilayer + NH$_3$	&   42	\\
\hline
 \end{tabular} 
\label{tab:gap}
\end{table} 

\begin{figure}
 \begin{tabular}{cc}
  a)\includegraphics[scale=0.3,bb=50 50 410 302]{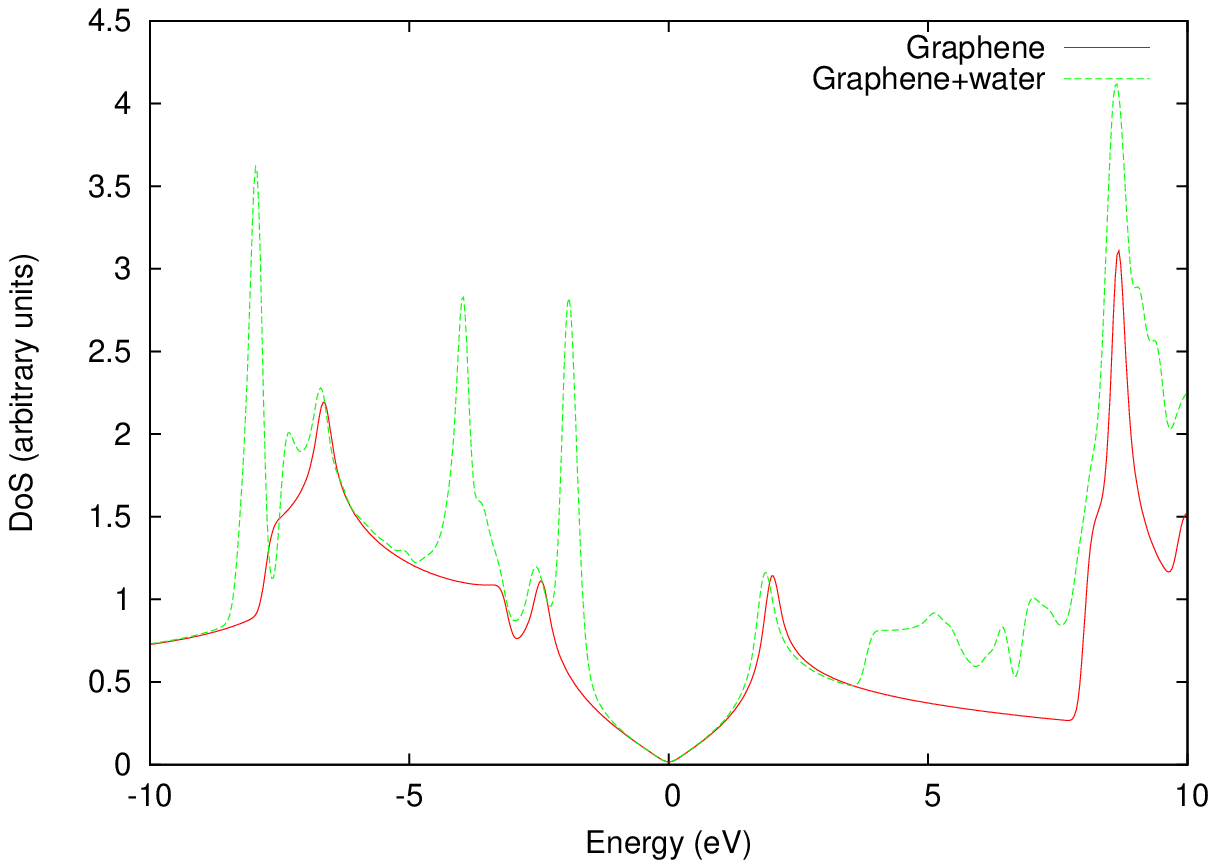}&
  c)\includegraphics[scale=0.3,bb=50 50 410 302]{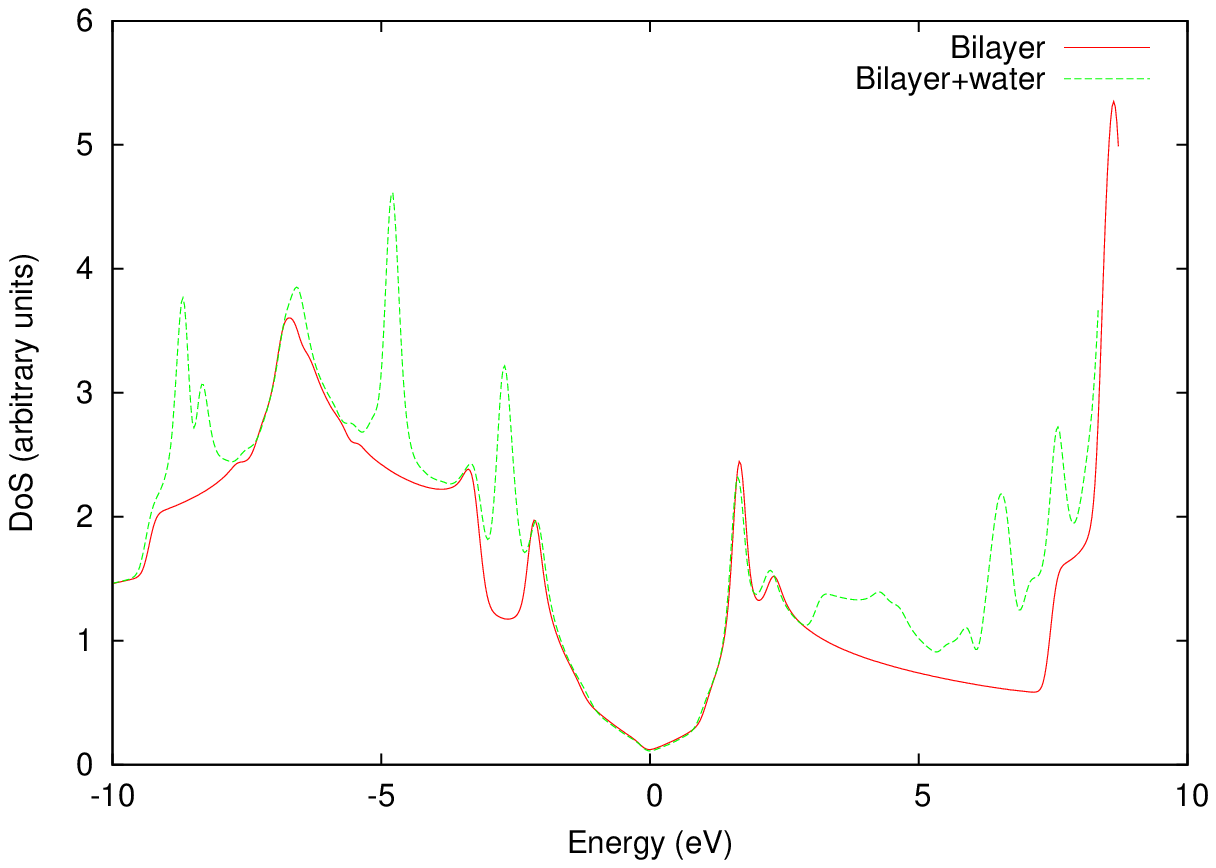}\\
  b)\includegraphics[scale=0.3,bb=50 50 410 302]{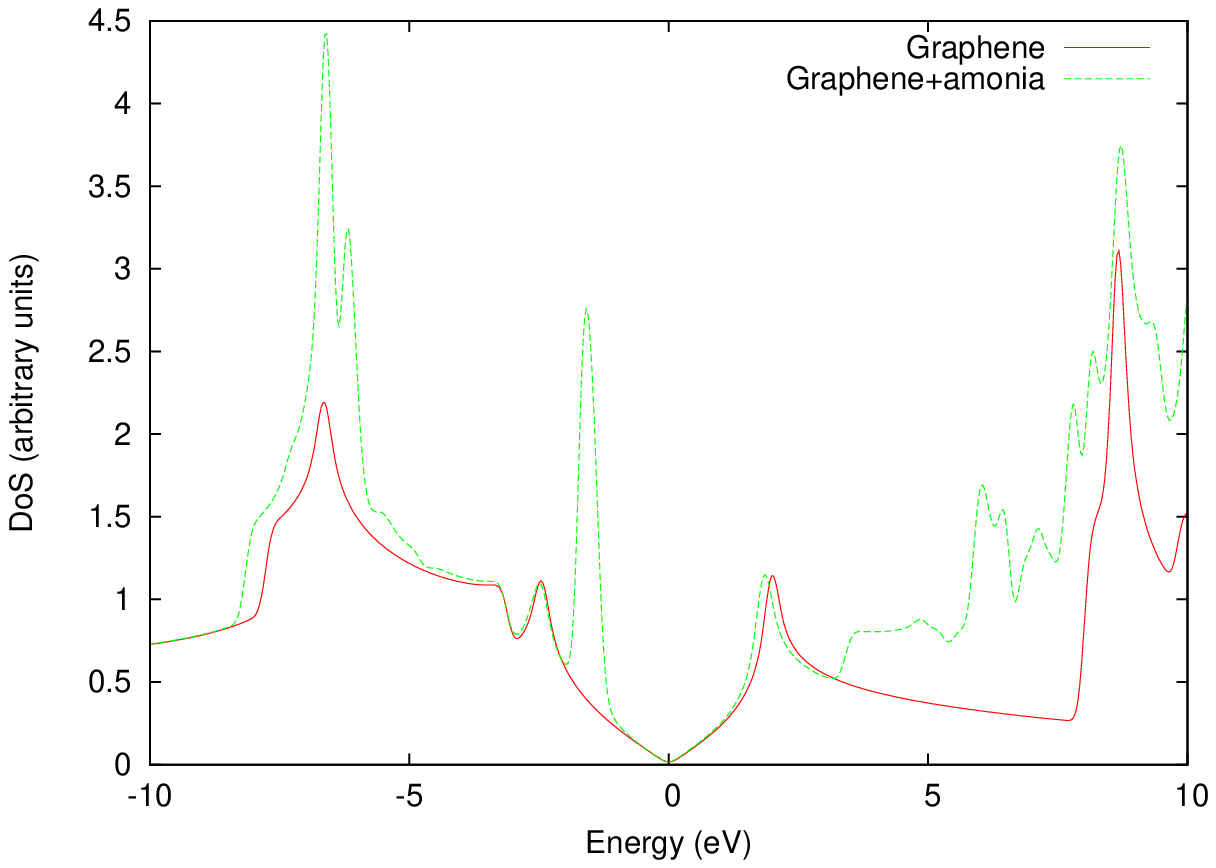}&
  d)\includegraphics[scale=0.3,bb=50 50 410 302]{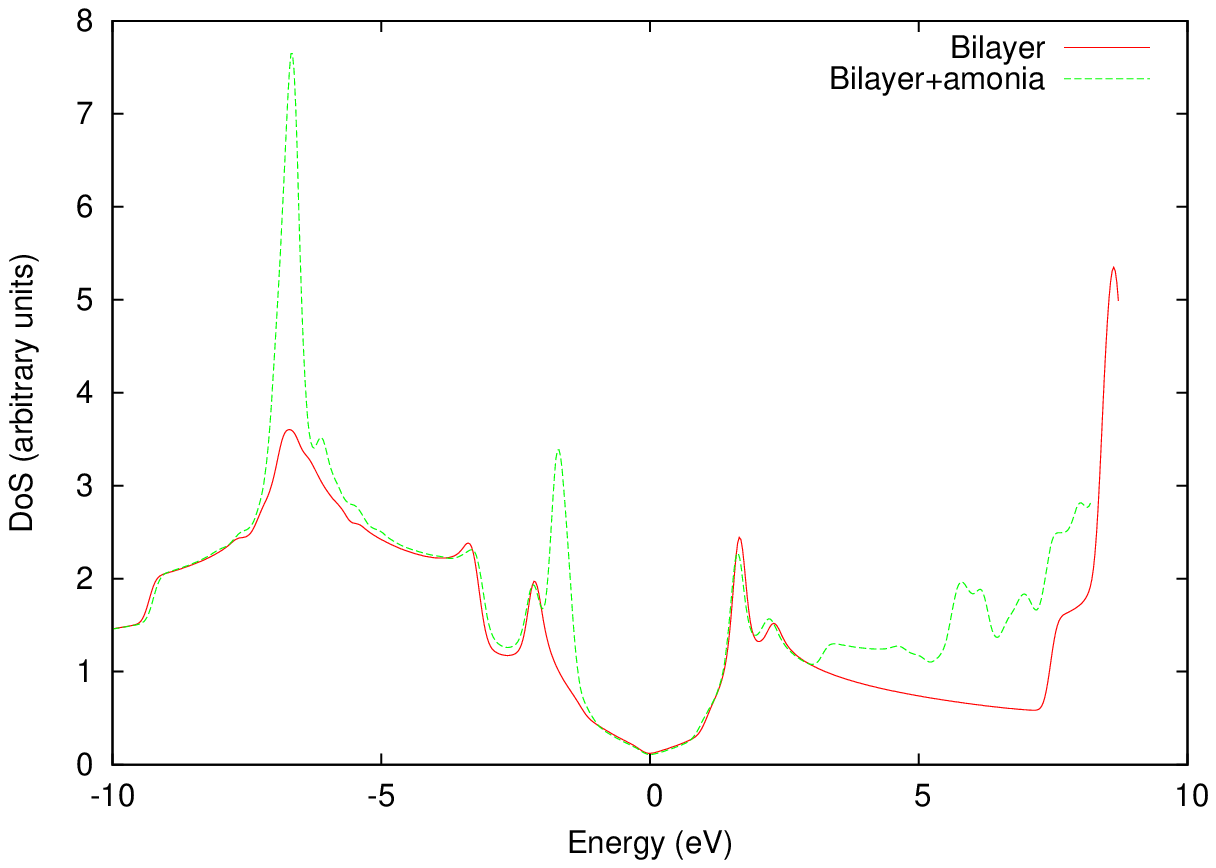}
 \end{tabular} 
\caption{(Color online) Electronic density of states for the system a) graphene plus water, b) graphene plus ammonia compared with graphene; c) bilayer plus water, d)bilayer plus ammonia compared with bilayer}
\label{fig:dos}
\end{figure}

The electronic density of states (DoS) of the systems considered is shown in figure \ref{fig:dos}.
The molecules create a lot of structure in the density of states, the most relevant are the strong peaks that correspond to very flat bands localized around the molecule.
Except for the water on the bilayer, the molecules generate a very high peak on the DoS at higher energies than the first peak of the occupied states for the clean surfaces.
The presence of these peaks leads to an increase of the absorption at energies of a few eV.

\section{Conclusions}

The adsorption of water and ammonia on the top of the graphene and graphene bilayer was studied.
The position and orientation of these molecules relative to the surfaces was obtained as well as the binding energies.
The high values of the binding energies (of the order of the {\ttfamily eV}) indicates that desorption of these molecules cannot be obtained at room temperature, and a surface coverage is expected if the surfaces are exposed to water or ammonia.

We have showed that
a gap  opens up in the band structure when molecules adhere to the surface, 
both on graphene and on the bilayer, and its value is of the order of a few tens of {\ttfamily meV}.
The mechanism for opening an energy gap may be different for the graphene single and bilayer, since
the random distribution of the molecules on the single layer may not break the A-B symmetry in average
while the gap-opening in the bilayer is related to the symmetry  between top and bottom layers.

We therefore have showed the existence of a feasible method to 
produce graphenes with gaps in their spectrum, which is not restricted to the
manipulation of nanoribbons.
We found that the  gap is in general displaced away from the Dirac point.

An increase in the absorption coefficient for energies above about 2~{\ttfamily eV} due to the presence of the molecules is predicted from the electronic density of states.

\begin{acknowledgments}
We wish to acknowledge the support of the Funda\c{c}\~ao para a Ci\^encia e a Tecnologia (FCT) under
the SeARCH (Services and Advanced Research Computing with HTC/HPC clusters) project, 
funded by FCT under contract CONC-REEQ/443/2005.
N.M.R.P. thanks the ESF Science Programme INSTANS
 2005-2010, and FCT under the grant PTDC/FIS/64404/2006.

\end{acknowledgments}



\end{document}